\documentclass[12pt, letterpaper]{JHEP3}

\usepackage{graphicx}
\usepackage{amssymb}

\def\gsim{\, \rlap{$>$}{\lower 1.1ex\hbox{$\sim$}}\,}
\def\lsim{\, \rlap{$<$}{\lower 1.1ex\hbox{$\sim$}}\,}

\title{Evaporation of large black holes in AdS: greybody factor and decay rate}

\author{
Jorge V. Rocha\\
Centro Multidisciplinar de Astrof\'isica - CENTRA, Dept. de F\'isica,\\
Instituto Superior T\'ecnico, Av. Rovisco Pais 1, 1049-001 Lisboa, Portugal\\ 
E-mail: \email{jorge.v.rocha@ist.utl.pt}
}

\abstract{
We consider a massless, minimally coupled scalar field propagating through the geometry of a black 3-brane in an asymptotically $AdS_5 \times S^5$ space.  The wave equation for modes traveling purely in the holographic direction reduces to a Heun equation and the corresponding greybody factor is obtained numerically.  Approximations valid in the low- and high-frequency regimes are also obtained analytically.  The greybody factor is then used to determine the rate of evaporation of these large black holes in the context of the evaporon model proposed in~\cite{Rocha:2008fe}.  This setting represents the evolution of a black hole under Hawking evaporation with a known CFT dual description and is therefore unitary.  Information must then be preserved under this evaporation process.
}

\begin{document}

\section{Introduction}

The AdS/CFT correspondence~\cite{Maldacena:1997re, Gubser:1998bc, Witten:1998qj} is a powerful conjecture which relates theories of gravity (and extensions) in spacetimes including an asymptotically anti-de Sitter (AdS) factor to conformal field theories (CFT) formally defined on the boundary of AdS.
Gravitational physics is completely absent in the explicit formulation of the CFT but is nevertheless somehow encoded therein by virtue of the duality.
In one of its manifestations, the correspondence states that the generating functional for the CFT is given by the partition function for the dual supergravity (SUGRA) theory, supplemented by certain boundary conditions for the SUGRA fields\footnote{See~\cite{Aharony:1999ti, D'Hoker:2002aw} for reviews on the AdS/CFT correspondence.}.
In this way each supergravity field is in one-to-one correspondence with a CFT operator: the non-normalizable solution for the SUGRA field plays the role of a source for the gauge invariant operator on the CFT side.
This state of affairs allows a dual interpretation of configurations (and its evolution) in the bulk of the spacetime in terms of a unitary field theory on the boundary.

One of the many applications of the gauge/gravity duality mentioned in the previous paragraph concerns the black hole information paradox~\cite{Hawking:1976ra}.
This is one of the major long-lasting conceptual problems in physics and it is related to the issue of unitarity in gravitational systems.
Long ago, Hawking showed by a semiclassical calculation that black holes radiate~\cite{Hawking:1974sw} and the resulting emission spectrum is that of a black body when measured near the event horizon.
The information paradox arises when one realizes that the complete evaporation by Hawking radiation of a black hole formed by the collapse of an initial pure state results in a mixed (thermal) state -- unitarity is violated in the process.
On the other hand, the manifest unitarity of the dual field theory indicates that information is preserved in the process of black hole formation and consequent evaporation~\cite{Lowe:1999pk}.

In an attempt to use AdS/CFT to address the information paradox it seems that we must consider black holes in asymptotically AdS spacetimes.
However, not all such black holes are good candidates.
The so-called {\em large} black holes in AdS are thermodynamically stable: their specific heat is positive~\cite{Hawking:1982dh}.
This means that they do not evaporate by Hawking emission.
Instead they reach a configuration of thermal equilibrium with the surrounding gas of particles.
Heuristically, the emitted radiation is reflected at the boundary and is reabsorbed by the black hole long before it evaporates.
Nevertheless, this class of black holes corresponds precisely to the situation which has a known dual description in terms of a high temperature thermal state in the CFT~\cite{Witten:1998zw}.
The class of {\em small} black holes in AdS has negative specific heat and is therefore thermodynamically unstable\footnote{See however Ref.~\cite{Asplund:2008xd} for some ideas regarding their dual interpretation.}.

The precise formulation of the AdS/CFT correspondence presented in~\cite{Gubser:1998bc, Witten:1998qj} requires imposing a Dirichlet boundary condition on the bulk fields\footnote{This condition can be relaxed, thus obtaining a dynamical metric for the boundary theory~\cite{Compere:2008us}.}.
This situation amounts to total reflectivity of the AdS boundary.
Therefore, we are forced to change the boundary conditions for the bulk fields in AdS in such a way that the boundary becomes partially absorbing, thus allowing the evaporation of a large black hole.
If this is attained, there is hope that the AdS/CFT prescription can be used to gain further insight into the information paradox.

In a previous paper~\cite{Rocha:2008fe} we constructed a model in which the above-mentioned total reflectivity of the AdS boundary is altered.
This was accomplished by gluing an additional $(1+1)$-dimensional flat spacetime to the boundary of ${\rm AdS}_5$ where only an auxiliary field $\sigma(t,z)$ (the evaporon) is allowed to propagate.
The evaporon interacts with the bulk scalar field $\Phi$ (hereafter referred to as the dilaton) only along the intersection of the two spaces.
If $t$ is the timelike coordinate, ${\bf x}$ are the spacelike boundary coordinates and $r$ represents the holographic coordinate of ${\rm AdS}_5$ then this hypersurface is defined by $r=\varepsilon$, where $\varepsilon$ is a regulator for the AdS bulk.
The dual interpretation is clear: we are adding an extra sector to the CFT and the desirable evaporation of a large black hole in AdS corresponds to the transfer of energy to this infinite external reservoir.

The interaction term in the action describing the dilaton-evaporon system couples the two fields in the simplest possible way, with a strength controlled by a constant $\lambda$:
\begin{equation}
S_{int}[\Phi,\sigma]  =   \lambda \int dt d^3x \, \sqrt{-h} \, \Phi(t,{\bf x},\varepsilon) \, \sigma(t,0)   \ ,
\label{S_int}
\end{equation}
where $h$ is the determinant of the metric induced on the interaction hypersurface.
The remainder of the action constitutes kinetic terms for the two fields plus a piece containing counterterms for $\sigma$: these are needed to renormalize the theory.
The counterterms were judicially computed in Ref.~\cite{Rocha:2008fe} where it was noted that their addition is crucial to obtain finite results when one takes the cutoff $\varepsilon \rightarrow 0$.

The main result of Ref.~\cite{Rocha:2008fe} is an expression for the transmission coefficient of a wave corresponding to a massless, minimally coupled scalar bulk field incident on the boundary:
\begin{equation}
|{\cal T}|^2(\omega)  =  \frac{2}{1 + \frac{1}{4\pi}\left(\frac{\omega_4}{\omega}\right)^3 + \pi\left(\frac{\omega}{\omega_4}\right)^3 \left[ 1 + \frac{4}{\pi^2} (\ln (\mu \, \omega) )^2 \right]} \ ,
\label{trans}
\end{equation}
where we have defined $\omega_4 \equiv  8 (2 \lambda^2 \kappa^2 R^5 V_3)^{-1/3}$. 
Here, $R$ is the AdS scale and $\kappa^2$ is related to the 5-dimensional Newton constant $G_5$ through $\kappa^2=8\pi G_5$.
The calculation that yielded this result was performed in Poincar\'e coordinates corresponding to the metric near the boundary of AdS.
In these coordinates the boundary is flat and infinite in extent.
Therefore to regulate the IR divergences, which would be absent if we considered AdS in global coordinates for which the boundary is ${\mathbb R}\times S^3$, we periodically identified the transverse space with a volume $V_3$ and $\mu$ is an infrared cutoff scale.

In principle the above framework allows black holes in AdS with known holographic duals to evaporate.
But to determine the rate of evaporation one ingredient is still missing: the greybody factor, which accounts for the fact that the emitted Hawking radiation needs to traverse frequency-dependent potential barriers due to the background geometry in order to reach asymptotic infinity.
The greybody factor thus corrects the otherwise perfect blackbody emission spectrum of a black hole.

There is by now a vast literature on this subject.
Early studies were conducted in the nineteen-seventies~\cite{Starobinskii:1973, Starobinskii:1974, Page:1976df, Unruh:1976fm, Sanchez:1977si}.
Greybody factors were also at the origin of the AdS/CFT correspondence, where they play a fundamental role in showing that the emission rates of black holes and of their corresponding D-brane systems precisely agree in various situations~\cite{Das:1996wn, Maldacena:1996ix}.
In the context of black holes in string theory, greybody factors have been obtained for several configurations, e.g.~\cite{Klebanov:1997cx, Klebanov:1997kc, Cvetic:1997uw, Gubser:1998iu}.
More recently, the idea of large extra dimensions and the prospect of producing black holes at the Large Hadron Collider has renovated the interest in the subject since the greybody factors can encode a fair amount of information pertaining to brane-world models (see Ref.~\cite{Kanti:2004nr} for a review, Refs.~\cite{Kanti:2005ja, Cardoso:2005mh, Jung:2007zz, Creek:2007sy, Casals:2008pq} and references therein for more recent studies in the same spirit).
In any event, greybody factors have been considered in a wide variety of cases: spherically symmetric, rotating and charged black holes and black branes in various dimensions, for several theories of gravity and particle spin.
Most relevant for our purposes, reference~\cite{Harmark:2007jy} presented analytic results valid in the low frequency regime for greybody factors of static, spherically symmetric black holes in asymptotically ${\rm AdS}_d$ spacetimes.
Nevertheless, there is a clear deficit of exact results in the literature regarding greybody factors in asymptotically AdS spacetimes.
In the present paper we shall present results for the black 3-brane in ${\rm AdS}_5$.

Given the setup of the evaporon model discussed above, and more specifically the fact that the computations were performed in Poincar\'e coordinates so that the boundary of ${\rm AdS}_5$ becomes ${\mathbb R} \times {\mathbb R}^3$, it is natural to consider a black hole possessing a horizon with three flat directions, i.e., a black 3-brane in ${\rm AdS}_5$.
The goal of the present paper is to compute the greybody factors for this geometry and consequently obtain the evaporation rate allowed in the framework of~\cite{Rocha:2008fe}.
We note that the analysis of~\cite{Harmark:2007jy}, even though concerned with spherically symmetric black holes, can be adapted to our case since the metric will only depend on one coordinate as well, namely the holographic coordinate.
Having obtained the greybody factor one can express the asymptotic flux of evaporons per unit frequency interval as the product of three terms: the blackbody spectrum for bosons $\left[ e^{\omega/T_H} -1 \right]^{-1}$, where $T_H$ is the temperature of the black brane\footnote{The Hawking radiation from black holes in AdS acquires the same blackbody form as in the asymptotically flat case~\cite{Hemming:2000as}.}, the greybody factor $\Gamma(\omega)$ and the transmission coefficient at the dilaton-evaporon interface, $|{\cal T}|^2$.
Correspondingly, the energy emitted per unit time is given by
\begin{equation}
\frac{dE}{dt}  =  \int \frac{d\omega}{2\pi} \frac{\omega \, \Gamma(\omega)}{e^{\omega/T_H} -1} |{\cal T}|^2(\omega)  \ .
\label{power_spec}
\end{equation}
Note that given the form of the interaction term~(\ref{S_int}) only the dilaton mode independent of the transverse directions ${\bf x}$ couples to the evaporon.
This corresponds to the planar limit of the s-wave.
Therefore the greybody factor that enters into the power spectrum~(\ref{power_spec}) is the one pertaining to the mode propagating purely in the holographic direction.
All other modes do not contribute.

The rest of the paper is organized as follows.
In section 2 we briefly describe the black 3-brane background for which we shall compute the greybody factors.
The formalism behind this calculation is presented in section 3, closely following Ref.~\cite{Harmark:2007jy}.
Here we also present the numerical results and compare them with the analytic predictions valid for low frequencies.
In the following section we use these results to obtain the decay rate of the black brane.
Section 5 contains the conclusions.
The appendices present derivations of the low- and high-frequency behavior of the greybody factor.

\section{The geometry and properties of the black 3-brane}

The metric which describes a black 3-brane with ${\rm AdS}_5 \times S^5$ asymptotics is the following~\cite{Horowitz:1998pq, Johnson:book}:
\begin{eqnarray}
ds^2  &=&  H^{-1/2}(r) \left[ - f(r) dt^2 + d{\bf x}^2 \right]  +  H^{1/2}(r) \left[ f^{-1}(r) dr^2 + r^2 d\Omega_5^2 \right]   \\
      &=&  - \frac{r^2}{R^2} \left( 1-\frac{r_H^4}{r^4}\right) dt^2 + \frac{r^2}{R^2} \, d{\bf x}^2 + \frac{R^2}{r^2} 
           \left(1-\frac{r_H^4}{r^4}\right)^{-1} dr^2 + R^2 d\Omega_5^2  \ ,  \nonumber
\end{eqnarray}
where
\begin{equation}
H(r) = \left( \frac{R}{r} \right)^4  \; \; \;  ,  \; \; \; \;  f(r) =  1 - \left( \frac{r_H}{r} \right)^4  \ .
\end{equation}
This is a solution of type IIB string theory in 10 dimensions of the form $X \times S^5$, where the five-sphere has constant radius $R$ and it corresponds to the decoupling limit of a stack of D3-branes.
The five dimensional manifold $X$ features a regular horizon at $r=r_H$ and in the limit $r \rightarrow \infty$ it approaches the metric of ${\rm AdS}_5$ in Poincar\'e coordinates,
\begin{equation}
ds^2_{AdS_5} =  \frac{r^2}{R^2} \left[ - dt^2 + d{\bf x}^2 \right]  +  \frac{R^2}{r^2} dr^2  \ .
\end{equation}
We see that the length $R$ also determines the AdS scale. 
The five-sphere plays no role in the following so we shall drop it from now on.

The Hawking temperature $T_H$ of the black 3-brane can be found by performing a Wick rotation and a change of the radial coordinate in such a way that the $(t,r)$-part of the metric becomes simply the metric of the Euclidean plane in polar coordinates.
To avoid a conical singularity at the origin the angular coordinate, which is proportional to the Euclidean time, must have periodicity $2\pi$ and this implies that
\begin{equation}
T_H  =  \frac{1}{4\pi} \frac{d}{dr} \left( \frac{f(r)}{\sqrt{H(r)}} \right)  =  \frac{r_H}{\pi R^2}  \ .
\label{T_H}
\end{equation}
The energy and entropy per unit 3-volume of this black hole are given by~\cite{Johnson:book, Mateos:2007vn}:
\begin{equation}
\frac{E}{V_3}  =  \frac{3 r_H^4}{2 \kappa^2 R^5}  \qquad \ , \qquad  \frac{S}{V_3} = \frac{2 \pi r_H^3}{\kappa^2 R^3} \ .
\label{BHmass}
\end{equation}
Observe that, since both the temperature and energy are proportional to positive powers of $r_H$, these black objects have positive specific heat and therefore do not evaporate if the boundary of AdS is totally reflective.
In this sense, these black branes fall in the category of the {\em large} black holes in AdS.

\section{The greybody factor}

\subsection{Formalism}

Consider a massless, minimally coupled scalar field $\Phi$ propagating in this geometry.
In particular, let this mode have definite frequency $\omega$ and wavevector ${\bf k}$,
\begin{equation}
\Phi(t,{\bf x},r) = e^{i \omega t - i {\bf k} \cdot {\bf x} } \phi_{\omega,{\bf k}}(r)  \ .
\end{equation}
One can always obtain the general solution as a sum over all possible modes.
The equation of motion for the scalar field is (from now on we suppress the subscripts $\omega,{\bf k}$)
\begin{equation}
\partial_r \left[ r^5 f(r) \partial_r \phi \right]  +  R^4 r \left[ f^{-1}(r) \omega^2 - {\bf k}^2 \right] \phi = 0  \ .
\label{EoM}
\end{equation}

Now let ${\bf k}^2 = m^2 \omega^2$.
After a coordinate change from $r$ to the dimensionless variable $u \equiv r_H^2/r^2$ the above equation of motion becomes
\begin{equation}
\frac{\partial^2 \phi}{\partial u^2}  -  \frac{1+u^2}{u(1-u^2)} \frac{\partial \phi}{\partial u}  +  \frac{\widetilde{\omega}^2}{4u(1-u^2)^2} \left( 1 - m^2(1-u^2) \right) \phi = 0  \ .
\label{Heun}
\end{equation}
According to the arguments presented at the end of the introduction, to obtain the rate of evaporation in the context of the evaporon model we just need to consider the case $m = 0$, i.e., modes traveling purely in the holographic direction.
But for now we keep $m$ general.
This will allow us to compute numerically (and analytically, in the low frequency regime) the greybody factor for a general value of $m$.
The absorption coefficient is expected to be lower for modes with $m \neq 0$, as this corresponds to a higher potential barrier that must be traversed to reach infinity (see eq.~(\ref{potential}) below).

We see that the problem has only a single parameter, namely the dimensionless combination of the frequency, the AdS scale and the horizon radius:
\begin{equation}
\widetilde{\omega} \equiv \omega \frac{R^2}{r_H}  \ .
\end{equation}
This is in line with the findings of~\cite{Horowitz:1999jd} where it was shown that for large AdS black holes the quasinormal frequencies must be proportional to the temperature.
Equation~(\ref{Heun}) is a second order linear ODE with 4 regular singular points, namely $u=0,1,-1,\infty$.
Such equations are known in the literature as Heun equations and complete solutions are beyond present knowledge.
Note that the region $r_H \leq r < \infty$ is mapped to the interval $u \in (0,1]$.

Instead of the holographic coordinate $r$ it is sometimes useful to consider the so-called tortoise coordinate $x$ which satisfies $dx = H^{1/2}(r) f^{-1}(r) dr$, or equivalently
\begin{equation}
\frac{dx}{dr} = \frac{R^2}{r^2} \left( 1 - \frac{r_H^4}{r^4} \right)^{-1}  \ .
\label{tortoise-diff}
\end{equation}
Upon integration the tortoise coordinate may be expressed in terms of $r$,
\begin{equation}
x(r) = \frac{R^2}{r_H} \left\{ \frac{1}{2} \arctan\left(\frac{r}{r_H}\right) + \frac{1}{4} \ln\left(8\frac{r-r_H}{r+r_H}\right) - \frac{\pi}{8} \right\}  \ .
\end{equation}
Equation~(\ref{tortoise-diff}) does not completely specify $x$.
We have chosen the integration constant in such a way that $x \simeq \frac{R^2}{4r_H} \log\left(4(r-r_H)/r_H\right)$ for $r-r_H \ll r_H$.
Approaching the horizon ($r \rightarrow r_H$) corresponds to the limit $x \rightarrow -\infty$.
The tortoise coordinate covers only the region outside the horizon.

In what comes next we shall follow the analysis of Ref.~\cite{Harmark:2007jy}.
The equation of motion for a scalar field mode with frequency $\omega$ can be recast in a Schr\"odinger-like form in one dimension:
\begin{equation}
\left[ \partial_x^2 + \omega^2 - V(r) \right] \left( r^{3/2} \phi \right) = 0  \ ,
\label{Schroed}
\end{equation}
where the potential $V$ is given by
\begin{equation}
V(r) = \frac{15}{4} \frac{r^2}{R^4} f(r)^2  +  6 \frac{r_H^4}{r^2 R^4} f(r) +  m^2\omega^2 f(r)  \ .
\label{potential}
\end{equation}
In the near-horizon region ($r \simeq r_H$ and $V(r) \ll \omega^2$) the solution is
\begin{equation}
\phi(r) = A e^{i \omega x}  \ .
\label{sol_hor}
\end{equation}
Here we chose the solution that is purely ingoing at the horizon.
Now, given the Schr\"odinger-like form of eq.~(\ref{Schroed}) it follows immediately that the current
\begin{equation}
j = \frac{1}{2i} \left( \phi^* \frac{d\phi}{dx} - \phi \frac{d\phi^*}{dx} \right)
\label{flux}
\end{equation}
is conserved.
This conserved current~(\ref{flux}) is nothing but the flux per unit {\em coordinate} area.
Thus, the flux per unit physical transverse area near the horizon is
\begin{equation}
J_{hor} = \frac{r_H^3}{R^3} \omega |A|^2  =  \frac{r_H^4}{R^5} \widetilde{\omega} |A|^2  \ .
\label{flux-hor}
\end{equation}

Next consider the equation of motion~(\ref{EoM}) in the asymptotic region.
For large $r$ (i.e. $r \gg r_H$) it reduces to
\begin{equation}
\left[ \frac{d^2}{dr^2} + \frac{5}{r} \frac{d}{dr} + (1-m^2)\omega^2 \frac{R^4}{r^4} \right] \phi(r) = 0  \ ,
\end{equation}
and the solutions may be expressed in terms of Hankel functions of degree two:
\begin{equation}
\phi(r) = C_1 (1-m^2)\frac{\omega^2 R^4}{r^2} H_2^{(1)}\left( \sqrt{1-m^2}\frac{\omega R^2}{r} \right)  +  C_2 (1-m^2)\frac{\omega^2 R^4}{r^2} H_2^{(2)}\left( \sqrt{1-m^2}\frac{\omega R^2}{r} \right)  \ .
\label{sol_asy}
\end{equation}
The first term is associated with the outgoing part of the wave-function $\phi$.
Similarly, the second term is identified with the ingoing piece.
If we now compute the asymptotic flux per unit {\em physical} area using eq.~(\ref{flux}) we obtain
\begin{equation}
J_{asy} = \frac{2}{\pi} (1-m^2)^2\omega^4 R^3 \left[ |C_2|^2 - |C_1|^2 \right]  =  \frac{2}{\pi} (1-m^2)^2\widetilde{\omega}^4 \frac{r_H^4}{R^5} \left[ |C_2|^2 - |C_1|^2 \right]  \ .
\end{equation}
The fact that the current~(\ref{flux}) is conserved manifests itself in the equality $J_{hor} = J_{asy}$.

To compute the greybody factor it is useful to define the ingoing and outgoing parts of the asymptotic flux:
\begin{equation}
J_{in} = \frac{2}{\pi} (1-m^2)^2\omega^4 R^3 |C_2|^2  \; \; \; \; \; \; ; \; \; \; \; \; \;  J_{out} = \frac{2}{\pi} (1-m^2)^2\omega^4 R^3 |C_1|^2  \ .
\label{flux-asy}
\end{equation}
In this case the greybody factor is defined~\cite{Harmark:2007jy} by the ratio of the (ingoing) flux at the horizon by the ingoing part of the asymptotic flux:
\begin{equation}
\Gamma_m(\omega) \equiv \frac{J_{hor}}{J_{in}}  \ .
\end{equation}
Given that the general solutions near the horizon and near the boundary are known, all we need is to find the interpolation that connects the two regions.
In other words, if the solution has the behavior displayed in eq.(\ref{sol_hor}) when $r \rightarrow r_H$, our task is to express the constant $C_1$ (and consequently $C_2$ as well) characterizing the asymptotic solution in terms of the constant $A$.
However, this program is frustrated by the lack of known global solutions to the Heun equation so we shall resort to numerical methods in the next section.
Nevertheless, an analytic expression can be obtained for the greybody factor in the limit of low frequencies (compared to the Hawking temperature of the black hole).
The analysis is similar to that performed in Ref.~\cite{Harmark:2007jy} (which concerned spherically symmetric black holes, but in our case the metric only depends on a single coordinate as well) and is relegated to Appendix A.
In this limit the greybody factor is given by
\begin{equation}
\Gamma_m(\omega) \simeq  \frac{\pi}{2} (1-m^2)^2 \left( \frac{R^2 \omega}{r_H} \right)^3 =  \frac{\pi}{2} (1-m^2)^2 \widetilde{\omega}^3  \ .
\label{GBF_m}
\end{equation}
This approximation is valid as long as the frequency is small compared to the temperature of the black brane $\omega \ll T_H$, or equivalently, $\widetilde{\omega} \ll 1/\pi$.
In the opposite limit, greybody factors at high energy (beyond the geometrical optics approximation) have been considered more recently in~\cite{Creek:2007sy}.
Appendix B contains a derivation of the greybody factor in the high frequency regime for the black branes considered in this paper.
This derivation resorts to the solution-matching technique as well, but in addition employs a WKB approximation to find the solutions away from the boundary of AdS.

\subsection{Numerical results}

We already mentioned in the previous section the overall strategy to compute the greybody factor.
Now we briefly describe the numerical implementation.
First, we choose to work with the coordinate $u$ because the physical region outside the horizon is converted to a finite interval in this coordinate, namely $(0,1]$.
The differential equation~(\ref{Heun}) is then solved numerically within this interval supplemented by a boundary condition imposed near the singular point $u=1$.
This boundary condition is nothing but the requirement that the wave is ingoing at the horizon, i.e. the numerical solution and its first derivative at $u=1-\epsilon$ are equal to eq.~(\ref{sol_hor}) and its first derivative, and then we let $\epsilon \rightarrow 0$.

To extract the coefficients $C_1$ and $C_2$ we first note that eq.~(\ref{sol_asy}) may be written in terms of Bessel functions as well.
Replacing $r$ by $r_H/\sqrt{u}$ we have
\begin{equation}
\phi(u) = (C_1+C_2) G_+(u)  +  i (C_1-C_2) G_-(u)  \ .
\end{equation}
with 
\begin{eqnarray}
G_+(u) &=& (1-m^2)\widetilde{\omega}^2 u J_2 \left( \widetilde{\omega} \sqrt{(1-m^2)u} \right)   \ , \\
G_-(u) &=& (1-m^2)\widetilde{\omega}^2 u Y_2 \left( \widetilde{\omega} \sqrt{(1-m^2)u} \right)  \ .
\end{eqnarray}
If we denote the numerical solution by $\phi_{NS}$ then the coefficients $C_1$ and $C_2$, for a given frequency $\omega$, can be obtained by
\begin{eqnarray}
(C_1-C_2) &=& {\lim}_{u \rightarrow 0} \frac{ \phi_{NS}(u)}{i G_-(u)}  \ , \\
(C_1+C_2) &=& {\lim}_{u \rightarrow 0} \frac{ \phi'_{NS}(u) - i (C_1-C_2)  G_-'(u) }{ G_+'(u) }  \ .
\end{eqnarray}
After the determination of these coefficients the greybody factor follows immediately from
\begin{equation}
\Gamma_m(\omega) = \frac{\pi}{2 \widetilde{\omega}^3 (1-m^2)^2} \frac{|A|^2}{|C_2|^2}  \ .
\end{equation}

\FIGURE[t]{
 \includegraphics[width=35pc, bb= 0pt 0pt 800pt 250pt]{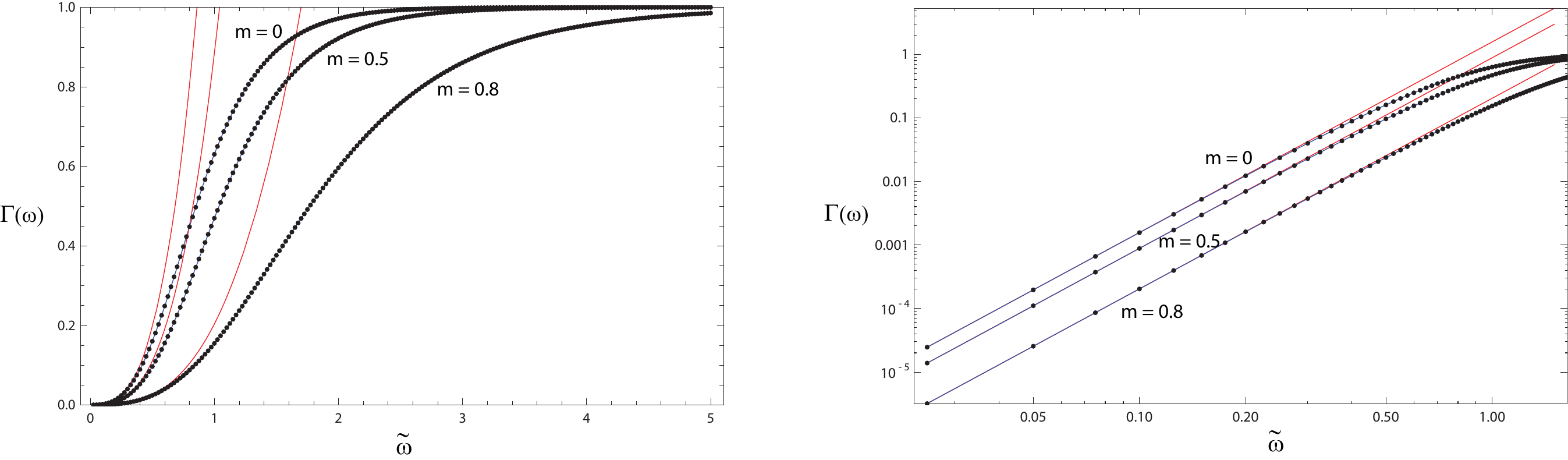} 
 \caption{The greybody factor is plotted as a function of the dimensionless frequency $\widetilde{\omega} = \omega R^2/r_H$ for three values of $m$.  The dots correspond to the numerical results and interpolating lines have been added to guide the eye.  The thin red lines represent to the analytic results valid in the limit of low frequencies.  On the right-hand panel a close-up around small $\widetilde{\omega}$ is displayed in a log-log plot.  The top curves correspond to $m=0$, the middle ones to $m=0.5$ and the lower ones to $m=0.8$.  The agreement between the numerical result and the analytic approximation at low frequencies is manifest.}
 \label{plot_gbf1}
}

The results are shown in Fig.~\ref{plot_gbf1} and Fig.~\ref{plot_gbf2}.
Figure~\ref{plot_gbf1} presents the greybody factors for various values of $m$.
We find that the greybody factor smoothly interpolates between $0$ and its asymptotic value $1$.
This is in line with the results of~\cite{Harmark:2007jy} who showed that, at least in the low-frequency regime, the large AdS black holes can never have $\Gamma(\omega) = 1$.
The numerical results are in very good agreement with the analytic approximation valid at low frequencies.
The greybody factor flattens out for larger $m$ and for $m=1$ it must vanish identically, as discussed in Appendix A.

\FIGURE[t]{
 \includegraphics[width=23pc, bb= 0pt 0pt 400pt 240pt]{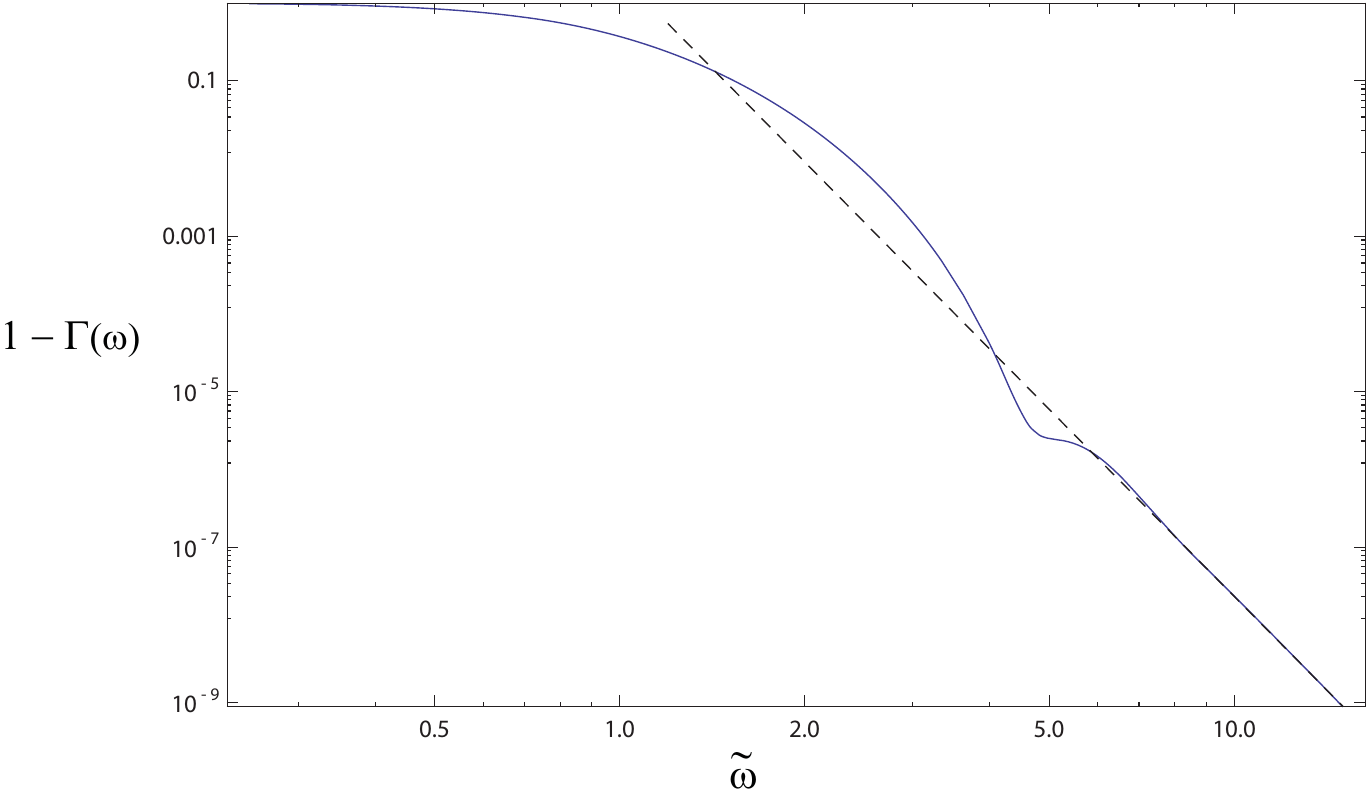} 
 \caption{The approach of the greybody factor to unity (its limiting value) is plotted as a function of the dimensionless frequency $\widetilde{\omega} = \omega R^2/r_H$ for $m=0$.  The dashed line corresponds to the fit $1-2.3\widetilde{\omega}^{-8}$ performed on the interval $\widetilde{\omega} \in \left[8,15\right]$.}
 \label{plot_gbf2}
}

Appendix B shows that the greybody factor tends to 1 at high frequencies.
The approach to this limiting value is also shown to be slower than $1/\widetilde{\omega}^3$.
This is confirmed by Fig.~\ref{plot_gbf2} where the approach of $\Gamma(\omega)$ to $1$ is seen to go like $\widetilde{\omega}^{-8}$ for large frequencies.
The small feature around $\widetilde{\omega} \sim 4.5$ persists when we increase the precision of the numerical calculations.
Its explanation remains unclear at the moment but one possibility is that it may be related to the lowest quasinormal mode of large AdS black holes.

\section{The decay rate}

In the previous section we obtained the greybody factor for the particular geometry under consideration.
This can now be used to compute the rate of evaporation of the black 3-branes by employing Eq.~(\ref{power_spec}).
Before proceeding, a few comments are in order.

First of all, we should note that the parameter $\omega_4$ that enters the transmission coefficient~(\ref{trans}) can be chosen freely by adjusting the coupling constant $\lambda$ of the evaporon model.
Secondly, given our ignorance about the IR cutoff $\mu$ we will integrate the power spectrum over frequencies small compared to $\omega_4$.
Indeed, for
\begin{equation}
\frac{\omega}{\omega_4}  \ll  \frac{1}{\sqrt[3]{2\pi}}   \ ,
\end{equation}
the log term in~(\ref{trans}) can be safely ignored.
This procedure yields a slight underestimate for the decay rate.
Introducing the notation $\widetilde{\omega}_4 = \omega_4 R^2 / r_H$ we then have
\begin{equation}
\frac{dE}{dt} \gsim  \frac{r_H^2}{2\pi R^4} \int_0^{\widetilde{\omega}_{max}} d\widetilde{\omega} 
\frac{\widetilde{\omega} \, \Gamma(\widetilde{\omega}) }{e^{\pi \widetilde{\omega}} -1}  \,
\frac{2}{1 + \frac{1}{4\pi} \left( \frac{\widetilde{\omega}_4}{\widetilde{\omega}} \right)^3 + \pi \left( \frac{\widetilde{\omega}}{\widetilde{\omega}_4} \right)^3  }    \ .
\label{rate:estimate}
\end{equation}
As discussed in the introduction, only the $m=0$ mode enters the computation of the decay rate.

The fist term in the above integral peaks\footnote{If the greybody factor were given by its low-frequency behavior~(\ref{GBF_m}) then the peak would be at $\widetilde{\omega} \simeq 1.25$.} roughly at $\widetilde{\omega} \approx 1$, whereas the maximum of the second term is determined by $\widetilde{\omega}_4$.
The rate of evaporation is basically given by the integral of the overlap of these two functions and a good estimate is obtained when the upper limit of integration satisfies
\begin{equation}
1  \ll  \widetilde{\omega}_{max}  \ll  \widetilde{\omega}_4    \ .
\end{equation}
Nevertheless, the rate is optimized by taking the smallest possible $\widetilde{\omega}_4$ for which the above condition can be met.
For example, choosing $\widetilde{\omega}_{max}=5$ and $\widetilde{\omega}_4=20$ the integral of Eq.~(\ref{rate:estimate}) gives roughly $2 \times 10^{-4}$.
Even though this procedure gives better estimates for the decay rate than what we shall consider below, it hides some of the dependence on the horizon radius in the parameter $\widetilde{\omega}_4$ and so it is difficult to follow the evaporation history of the black brane.

If on the other hand we take $\widetilde{\omega}_{max} \ll 1/\pi$ we can use the low-frequency approximation for the greybody factor and obtain analytically an underestimate for the decay rate, which however yields the exact dependence on $r_H$.
Using Eq.~(\ref{GBF_m}) and noting that the second term in the denominator of the transmission coefficient largely dominates the others we arrive at
\begin{equation}
\frac{dE}{dt}  >  \frac{\pi r_H^5}{\omega_4^3 R^{10}} \int_0^{\widetilde{\omega}_{max}} d\widetilde{\omega} 
\frac{\widetilde{\omega}^7 }{e^{\pi \widetilde{\omega}} -1}   \ .
\label{rate:lowbound}
\end{equation}
This integral can be computed exacly and it is given in terms of polylogarithms.
For $\widetilde{\omega}_{max} \simeq 0.1$ its numerical value is $4 \times 10^{-9}$.
For the same choice as above, $\widetilde{\omega}_4 = 20$, the lower bound~(\ref{rate:lowbound}) is roughly 8 orders of magnitude smaller than that obtained from the estimate~(\ref{rate:estimate}).
In any event, the lower bound~(\ref{rate:lowbound}) allows us to obtain an upper limit for the evaporation time.
Using the relation between the mass of the black brane and its horizon radius, Eq.~(\ref{BHmass}) with $V_3 = {\rm Vol}(S^3) R^3 = 2\pi^2 R^3$,  we find that the time it takes for the horizon length to drop from $r_H^{ini}$ to $r_H^{fin}$ is bounded by
\begin{equation}
\Delta t < 10^9 \frac{3\pi \omega_4^3 R^8}{\kappa^2} \left( \frac{1}{r_H^{fin}} - \frac{1}{r_H^{ini}} \right)   \ .
\end{equation}
For illustration purposes, taking the initial configuration to have an infinitely large horizon radius, the time elapsed until the black hole reaches the category of `small black holes', i.e., when $r_H^{fin} \sim R$, is less than $10^9 \frac{3\pi \omega_4^3 R^7}{\kappa^2}$.
An improved bound by several orders of magnitude is achievable by integrating numerically the rate of evaporation~(\ref{rate:estimate})  but this involves specifying the parameter $\omega_4$.
A full account of the black brane evaporation history is outside the scope of this paper.

In any case, the evaporon model allows the large black holes in AdS to evaporate in a finite time.
According to~\cite{Page:1993wv} information must start being recovered in the Hawking particles as soon as a black hole radiates half its initial entropy if the evolution is unitary.
Given the relation~(\ref{BHmass}) this will occur as long as $r_H^{ini} \geq \sqrt[3]{2} r_H^{fin}$.
Therefore, for large enough initial horizon length one still recovers the black hole information paradox and it is conceivable that the evaporon model can be used to shed some light on that long-standing issue.

\section{Conclusions}

The AdS/CFT correspondence provides a promising playground to investigate the black hole information paradox as it relates a gravitational theory to a unitary quantum field theory.
This has been apparent from the early days of the gauge/gravity duality but an explicit framework in which black holes with known CFT duals are allowed to evaporate is not easy to find.
The model developed in~\cite{Rocha:2008fe} permits such a situation by coupling a bulk field to the evaporon, a scalar field living on an auxiliary flat space.
This corresponds to coupling an extra sector to the CFT, thus providing an infinite reservoir for the field theory to cool down.

To compute the evaporation rate of the so-called large black holes in AdS, which is made possible by the evaporon model, one needs the corresponding greybody factor.
This is just the probability for a wave emitted from the black hole horizon to be transmitted though the potential barrier created by the spacetime geometry and reach infinity.
In this work we have studied this transmission coefficient for a minimally coupled, massless scalar field on a black 3-brane background with $AdS_5$ asymptotics.
This geometry falls in the category of `large' AdS black holes.
We computed the greybody factor both analytically in the low- and high-frequency limits and numerically, having obtained excellent agreement in the frequency regimes where comparison is meaningful.

The explicit results for the greybody factor allowed us to obtain an estimate and a lower bound on the evaporation rate of the black brane.
The evaporon model thus permits a large black hole in AdS to evaporate in finite time until it becomes a small AdS black hole.
During this whole evolution most of the initial entropy of the black hole is radiated (for large enough initial black holes) and the dual description is controlled, indicating a unitary process.

\acknowledgments

I wish to thank Joseph Polchinski for the suggestion and discussions that were at the origin of this project, and Vitor Cardoso for reading a final draft of this paper.  I have also benefited from conversations with Mariam Bouhmadi-L\'opez, Vitor Cardoso, Marc Casals and Paolo Pani.  This work was partially supported by Funda\c{c}\~ao para a Ci\^encia e Tecnologia (FCT) - Portugal through project PTDC/FIS/64175/2006.  I also acknowledge financial support from FCT fellowship SFRH/BPD/47332/2008.

\begin{appendix}
\renewcommand{\theequation}{A.\arabic{equation}}
  \setcounter{equation}{0}  
  \section{Low-frequency behavior of the greybody factor}  

This appendix reproduces the calculation of the greybody factor for low frequencies compared to the Hawking temperature of the black 3-brane in the ${\rm AdS}_5$ background.
It is a very slight generalization of the study performed in Ref.~\cite{Harmark:2007jy} using the matching technique, in the sense that we are able to consider modes with $m \neq 0$, i.e. not purely transverse to the brane, whereas the computations of Ref.~\cite{Harmark:2007jy} were only done for the $\ell = 0$ mode.
However, we specialize our discussion to $d=5$ spacetime dimensions.

We are interested in the low frequency regime, $\omega \ll T_H$, or equivalently, using Eq.~(\ref{T_H})
\begin{equation}
\widetilde{\omega} \ll 1/\pi  \ .
\end{equation}
Under this assumption it is not hard to see that we must have $r \simeq r_H$ in the region defined by $V(r) \ll \omega^2$.
This is the near-horizon region and the corresponding incoming solution was given in Eq.~(\ref{sol_hor}).
As in Ref.~\cite{Harmark:2007jy} we can move slightly away from the horizon while still remaining in the near-horizon region and the solution becomes
\begin{equation}
\phi(r) \simeq A \left[ 1 + \frac{i}{4}\widetilde{\omega} \log\left(4\frac{r-r_H}{r_H}\right) \right]  \ .
\label{sol_near-hor}
\end{equation}

Consider now the intermediate region where $V(r) \gg \omega^2$.
This also implies that $V(r) \gg \omega^2(1-m^2 f(r))$ and therefore the equation of motion in this region becomes $\partial_r\left[r^5 f(r) \partial_r \phi \right] = 0$.
The general solution is
\begin{equation}
\phi(r) = B_1 + \frac{B_2}{4r_H^4} \log\left(1-\frac{r_H^4}{r^4}\right)  \ .
\label{sol_inter}
\end{equation}
Matching this with~(\ref{sol_near-hor}) one finds
\begin{equation}
B_1 = A  \; \; \; \; \; \;   ;   \; \; \; \; \; \;  B_2 = A i \widetilde{\omega} r_H^4   \ .
\end{equation}
It is easy to check that in the low frequency regime the asymptotic region ($r \gg r_H$) is contained in the intermediate region (defined by $V(r) \gg \omega^2$).
Hence, we can match the wave-function~(\ref{sol_inter}) onto the solution for the asymptotic region~(\ref{sol_asy}) by considering the limit $r \gg r_H$.
The former and latter expressions become respectively
\begin{eqnarray}
\phi(r)  &\simeq&  A \left( 1 - \frac{i\widetilde{\omega}r_H^4}{4r^4} \right)   \ ,  \\
\phi(r)  &\simeq&  (C_2-C_1)\frac{4i}{\pi}  +  (C_1+C_2)(1-m^2)^2 \frac{\widetilde{\omega}r_H^4}{8r^4}   \ .
\end{eqnarray}
Thus, we find that
\begin{eqnarray}
C_1  &=&  -\frac{iA}{(1-m^2)^2} \left[ \frac{1}{\widetilde{\omega}^3} - (1-m^2)^2 \frac{\pi}{8} \right]   \ ,  \\
C_2  &=&  -\frac{iA}{(1-m^2)^2} \left[ \frac{1}{\widetilde{\omega}^3} + (1-m^2)^2 \frac{\pi}{8} \right]   \ .
\end{eqnarray}
Plugging the expression for $C_2$ into Eq.~(\ref{flux-asy}) we obtain an expression for the greybody factor for the $m^{\rm th}$ mode of a scalar field propagating in the geometry of a black 3-brane with ${\rm AdS}_5$ asymptotics, valid for low frequencies:
\begin{equation}
\Gamma_m(\omega) = \frac{J_{hor}}{J_{in}} = \frac{\pi}{2} (1-m^2)^2 \widetilde{\omega}^3   \ .
\end{equation}
Note that the greybody factor vanishes for $m = 1$ as expected since this corresponds to propagation parallel to the brane.

\renewcommand{\theequation}{B.\arabic{equation}}
  \setcounter{equation}{0}  
  \section{Large-frequency behavior of the greybody factor}  

In this appendix we derive the large-frequency behavior of the greybody factor using the WKB approximation.
For this purpose it is useful to recast Eq.~(\ref{Heun}) in the following form:
\begin{equation}
\frac{\partial^2 \phi}{\partial z^2}  +  \frac{\widetilde{\omega}^2}{16} \frac{1 - m^2 e^{-z}}{(1-e^{-z})^{3/2}} \phi = 0  \ .
\label{WKB}
\end{equation}
This is accomplished by an appropriate change of variables, $z = - \ln(1-u^2)$.
We are interested in obtaining an approximate solution of Eq.~(\ref{WKB}) valid for large $\widetilde{\omega}$.
Thus we write
\begin{equation}
\phi(z) =  e^{i \widetilde{\omega} f(z)}  \qquad , \qquad
f(z)    =  \sum_{n=0}^\infty \widetilde{\omega}^{-n} f_n(z)   \ ,   \label{WKBexpansion}
\end{equation}
and use $\epsilon = 1/\widetilde{\omega}$ as a small parameter for the perturbative expansion.
Eq.~(\ref{WKB}) then becomes
\begin{equation}
\frac{i}{\widetilde{\omega}}f''(z) - f'(z)^2 + \frac{1-m^2 e^{-z}}{16(1-e^{-z})^{3/2}}  = 0  \ .
\end{equation}
Inserting expansion~(\ref{WKBexpansion}) and working order by order in powers of $1/ \widetilde{\omega}$ one obtains
\begin{eqnarray}
f_0'(z) &=& \pm \frac{\sqrt{1-m^2 e^{-z}}}{4(1-e^{-z})^{3/4}}  \ , \nonumber \\
f_1'(z) &=& \frac{i f_0''(z)}{2 f_0'(z)} = - \frac{i}{8} e^{-z} \frac{3(1-m^2) + m^2(1-e^{-z})}{(1-e^{-z})(1-m^2 e^{-z})}  \ , \\
f_2'(z) &=& \frac{i f_1''(z) - f_1'(z)^2}{2 f_0'(z)}   \nonumber \\
        &=& - e^{-z} \frac{8(2m^2-3) + (9+4m^2+4m^4)e^{-z} + 10m^2(1-2m^2)e^{-2z} + m^4e^{-3z}}{32(1-e^{-z})^{5/4}(1-m^2e^{-z})^{5/2}}  \nonumber \ .
\end{eqnarray}
The integration of the above functions can be performed analytically in the case $m=0$~\footnote{The integration can also be done when $m=1$ but this corresponds to the scalar field propagating parallel to the black 3-brane.}, to which we now restrict ourselves.
Upon integration we get
\begin{eqnarray}
f_0(z) &=& \pm \frac{\pi}{4}(1-i) \left[ 1 - \frac{1}{\sqrt{2}\pi} B_{e^z}\left( \frac{3}{4},\frac{1}{4}\right) \right]   \ , \\
f_1(z) &=& - \frac{3 i}{8} \ln (1-e^{-z})   \ ,  \\
f_2(z) &=& - \frac{12+3e^{-z}}{8(1-e^{-z})^{1/4}}   \ ,
\end{eqnarray}
where $B_w(a,b)$ is the incomplete beta function.
Thus, to first order in the WKB approximation, the solution to Eq.~(\ref{WKB}) which is purely incoming at the horizon ($z\rightarrow\infty$) is
\begin{equation}
\phi_{WKB}(z) =  b (1-e^{-z})^{3/8}  \exp \left\{ -i \widetilde{\omega} \frac{\pi}{4}(1-i) \left[ 1 - \frac{1}{\sqrt{2}\pi} B_{e^z}\left( \frac{3}{4},\frac{1}{4}\right) \right] \right\}  \ .
\label{WKB_sol}
\end{equation}
This solution is valid for
\begin{equation}
\widetilde{\omega} \gg \frac{|f_0''|}{|f_0'|^2} = \frac{3e^{-z}}{(1-e^{-z})^{1/4}}   \ .
\end{equation}

The solution near the boundary ($z \simeq 0$) is given in Eq.~(\ref{sol_asy}).
We may express it in terms of the coordinate $z$ as
\begin{equation}
\phi_{NB}(z) = \alpha (\widetilde{\omega}z^{1/4})^2 J_2(\widetilde{\omega}z^{1/4}) 
        + \beta (\widetilde{\omega}z^{1/4})^2 Y_2(\widetilde{\omega}z^{1/4})   \ .
\label{sol_asy_z}
\end{equation}
This must be matched onto the WKB solution to determine the relation between the constants $\alpha$, $\beta$ and $b$.
To this end we introduce the parameter $\nu$ through
\begin{equation}
\widetilde{\omega} = \nu (z^{-1/4} + 1)   \ .
\end{equation}
For large $\nu$, but still in the limit $z \rightarrow 0$ and therefore for large $\widetilde{\omega}$, Eq.~(\ref{sol_asy_z}) becomes
\begin{equation}
\phi_{NB} \simeq  \nu^{3/2} \cos \nu \left[ \frac{\beta-\alpha}{\sqrt{\pi}} + O(\nu^{-1}) \right]  
          + \nu^{3/2} \sin \nu \left[ - \frac{\alpha+\beta}{\sqrt{\pi}} + O(\nu^{-1}) \right]  \ .
\end{equation}
On the other hand, if we take the $z \rightarrow 0$ limit on the WKB solution~(\ref{WKB_sol}) keeping $\nu$ fixed and large\footnote{This ensures the validity of the WKB approximation.}, we obtain
\begin{equation}
\phi_{WKB} \simeq  \frac{b}{\widetilde{\omega}^{3/2}} \nu^{3/2} e^{-i \nu}   \ .
\end{equation}
Matching the two solutions $\phi_{NB}$ and $\phi_{WKB}$ we find that $\alpha = \frac{\sqrt{\pi}}{2\widetilde{\omega}^{3/2}} (i-1) b$ and $\beta = \frac{\sqrt{\pi}}{2\widetilde{\omega}^{3/2}} (1+i) b$.
In terms of the solution presented in Eq.~(\ref{sol_asy}) we have 
\begin{equation}
C_2 = \frac{\sqrt{\pi}}{2\widetilde{\omega}^{3/2}} (i-1) b   \ .
\label{C2b}
\end{equation}

Using Eq.~(\ref{flux}) to compute the conserved current at the horizon ($e^{-z} = 0$) one obtains
\begin{equation}
J_{hor} =  \frac{r_H^4}{R^5} \widetilde{\omega} |b|^2   \ .
\end{equation}
This turns out to be equal to the incoming part of the asymptotic conserved current, given the relation~(\ref{C2b}).
Therefore we obtain that the greybody factor asymptotes to $1$ at large real frequencies, as expected.
One can consider higher order corrections to the greybody factor by including higher orders in the WKB expansion in $\widetilde{\omega}^{-1}$, and therefore hope to obtain the rate at which it approaches the limiting value $1$.
However, it appears that up to $4^{th}$ order all the corrections vanish and at $5^{th}$ order the validity of the WKB approximation breaks down.

\end{appendix}


\end{document}